\documentclass[letterpaper,11pt]{article}

\usepackage{amsmath,amssymb,epsfig,bbm}
\usepackage[active]{srcltx}
\usepackage[all]{xypic}

\usepackage{color}

\usepackage{dsfont}

\definecolor{co}{cmyk}{0,0.7,0.3,0}
\definecolor{darkgreen}{cmyk}{1,0,1,.2}
\definecolor{m}{rgb}{1,0.1,1}

\newcommand{\be}{\begin{equation}}
\newcommand{\ba}{\begin{eqnarray}}
\newcommand{\ea}{\end{eqnarray}}
\newcommand{\nn}{\nonumber}

\def\a{\alpha}
\def\b{\beta}

\def\e{\epsilon}

\def\l{\lambda}

\def\t{\tau}

\def\x{\xi}

\def\G{\Gamma}

\def\OO{\Omega}

\def\S{\Sigma}

\def\X{\Xi}

\def\ca{{\cal A}}
\def\cb{{\cal B}}

\def\ce{{\cal E}}
\def\cf{{\cal F}}

\def\ch{{\cal H}}

\def\ck{{\cal K}}
\def\cl{{\cal L}}
\def\cm{{\cal M}}
\def\cn{{\cal N}}

\def\cp{{\cal P}}

\newcommand{\pa}{\partial}

\newtheorem{thm}{Theorem}[subsection]

\newtheorem{conj}[thm]{Conjecture}
\newtheorem{definition}[thm]{Definition}

\newcommand{\bbC}{{\Bbb C}}
\newcommand{\bbR}{{\Bbb R}}
\newcommand{\cF}{{\cal F}}

\newcommand{\Tr}{\operatorname{Tr}}

\fontfamily{yfrak}

\begin{document}

\vskip 25mm

\begin{center}

{\Large\bfseries 
$ C^*$-algebras of Holonomy-Diffeomorphisms   \\[1ex]\& Quantum Gravity
 I 
}

\vskip 4ex

Johannes \textsc{Aastrup}$\,^{a}$\footnote{email: \texttt{aastrup@math.uni-hannover.de}} \&
Jesper M\o ller \textsc{Grimstrup}\,$^{b}$\footnote{email: \texttt{jesper.grimstrup@gmail.com}}\\ 
\vskip 3ex  

$^{a}\,$\textit{Institut f\"ur Analysis, Leibniz Universit\"at Hannover, \\ Welfengarten 1, 
D-30167 Hannover, Germany.}
\\[3ex]
$^{b}\,$\textit{Wildersgade 49b, 1408 Copenhagen K, Denmark.}\\[3ex]

\end{center}

\vskip 3ex

\begin{abstract}

A new approach to a unified theory of quantum gravity based on noncommutative geometry and canonical quantum gravity is presented. The approach is built around a $*$-algebra generated by local holonomy-diffeomorphisms on a 3-manifold and a quantized Dirac type operator; the two capturing the kinematics of quantum gravity formulated in terms of Ashtekar variables.  
We prove that the separable part of the spectrum of the 
algebra is contained in  the space of measurable connections modulo gauge transformations and we give  limitations to the non-separable part. 
The construction of the Dirac type operator -- and thus the application of noncommutative geometry -- is motivated by the requirement of diffeomorphism invariance. We conjecture that a semi-finite spectral triple, which is invariant under volume-preserving diffeomorphisms, arise from a GNS construction of a semi-classical state. 
Key elements of quantum field theory emerge from the construction in a semi-classical limit, as does an almost commutative algebra. 
Finally, we note that the spectrum of loop quantum gravity emerges from a discretization of our construction. Certain convergence issues are left unresolved. 

This paper is the first of two where the second paper \cite{AGnew} is concerned with mathematical details and proofs concerning the spectrum of the holonomy-diffeomorphism algebra.

\end{abstract}

\newpage
\tableofcontents
\newpage

\section{Introduction}

A new approach to a unified theory of quantum gravity, which combines elements of canonical quantum gravity with noncommutative geometry, is presented. This approach is built around two quantum variables: An algebra, which encodes the {\it holonomy-diffeomorphisms} of a 3-dimensional spin manifold, and  a Dirac type operator, which is at least invariant under volume-preserving diffeomorphisms.
These variables, which capture the kinematics of quantum gravity, are organized in a spectral triple construction. This application of noncommutative geometry is motivated by the requirement of diffeomorphism invariance.\\

The cornerstone of this approach is
the $C*$-algebra $\mathbf{HD}(M) $  generated by {holonomy diffeomorphisms}. These are lifts of local flows in a three dimensional manifold $M$ to a spin-bundle and are path dependent. 
%
 We prove that the separable part of the spectrum  of this noncommutative algebra  --  which is defined to be the set of its irreducible representations modulo unitary equivalence  --  is contained in the space of measurable connections modulo gauge equivalence. Furthermore, we find certain restrictions on the non-separable part of the spectrum. These results invite an interpretation of $\mathbf{HD}(M) $ as an algebra over a configuration space of Ashtekar connections  \cite{Ashtekar:1986yd,Ashtekar:1987gu} and thus establishes a connection  to canonical quantum gravity in $3+1$ dimensions. 

The definition of $\mathbf{HD}(M) $ is manifestly coordinate independent. However, in order to construct operators, which correspond to the canonical conjugate of the Ashtekar connection  --  these are densitised triad fields  --  , we need to introduce a coordinate-dependent formulation of $\mathbf{HD}(M) $. This formulation relies on an infinite sequence of nested cubic lattices, the totality of which amounts to a coordinate system. In this picture an element  $F$ in $\mathbf{HD}(M) $ is given by an infinite sequence of increasingly good approximations of $F$, each associated to a finite graph. At the level of a finite graph the configuration space of connections is given by various copies of the gauge group, and an element in $\mathbf{HD}(M) $ is approximated by a number of parallel transports in the graph.
The densitised triad operators are then given by infinite sequences of derivations on various copies of the gauge group associated to graphs.
The interaction between $\mathbf{HD}(M) $ and the densitized triad operators captures the kinematics of quantum gravity.

In a similar manner we also construct semi-classical states on $\mathbf{HD}(M) $. These states are infinite sequences of states associated to finite graphs and are essentially identical to the states constructed in \cite{AGNP1}-\cite{Aastrup:2011dt}. At the level of a finite graph they are given by coherent states on the various copies of the gauge group. We consider the GNS construction of a semi-classical state and obtain what amounts to a kinematical Hilbert space. This means that in this approach
each semi-classical approximation entails a {\it different} kinematical Hilbert space. Thus, we propose a construction where there is no universal kinematical Hilbert space. Certain convergence issues concerning the coherent states are not addressed.

The introduction of a coordinate system raises the question whether the construction is background dependent. 
Since the construction is built around the algebra $\mathbf{HD}(M) $ which includes the diffeomorphisms of $M$, there is an action of the diffeomorphism group in any representation of $\mathbf{HD}(M) $.
It turns out that a Dirac type operator commutes with -- at least --  the volume-preserving diffeomorphisms on $M$. This operator is an infinite sequence of Dirac operators associated to finite graphs and is essential identical to the operator analyzed in \cite{AGNP1}-\cite{AGN3}.
Thus, we find that the issue of diffeomorphism invariance motivates the construction of a Dirac type operator and -- bearing in mind that $\mathbf{HD}(M) $ is noncommutative -- lands us deep within the territory of noncommutative geometry.

The constructions presented in this paper are very similar to the constructions analyzed in \cite{AGNP1}-\cite{Aastrup:2011dt} (see \cite{Aastrup:2012jj} for a review and \cite{Aastrup:2005yk}-\cite{Aastrup:2009ux} for earlier versions of the framework). There, a spectral triple construction based on a projective system of graphs was constructed and it was shown that key elements of fermionic quantum field theory  -- the Dirac Hamiltonian and many-particle states -- emerge in a semi-classical approximation. 
The present construction differs from that of its predecessors on important issues, most notably the continuum limit, but key elements -- the Dirac type operator, the semi-classical states -- are kept essentially unaltered.
This means that the results in \cite{AGNP1}-\cite{AGN3} on the emergence of fermionic quantum field theory also apply to the setup presented here.

In a semi-classical approximation the holonomy-diffeomorphism algebra reduces to a semi-direct product between an almost commutative algebra and the diffeomorphism group. In the same limit the Hilbert space reduces to an $L^2$-space over $M$ and the Dirac type operator gives a spatial Dirac operator. All together we see the contour of an almost commutative geometry emerge in a semi-classical approximation. Albeit still on an tentative level this opens a door to a comparison with Connes' work on the standard model of particle physics, in which the entire standard model coupled to general relativity is formulated as a single gravitational model in terms of a spectral triple over a certain almost commutative algebra.\\

In technical terms we find the following: Associated to an infinite sequence of finite, nested graphs there exist an infinite sequence of Hilbert $C^*$-modules over algebras generated by local volume-preserving diffeomorphisms restricted to these graphs. This sequence of Hilbert $C^*$-modules carries an action of $\mathbf{HD}(M) $. The Dirac type operator acts in these Hilbert $C^*$-modules and gives rise to an infinite sequence
of unbounded Kasparov modules with a left-action by a certain subalgebra of the commutant of the volume-preserving diffeomorphism algebra. 
There is, at a finite level, a natural trace over the approximations of the diffeomorphism group, which promotes the Kasparov modules to semi-finite spectral triples. At the present level of analysis we leave certain important questions regarding the convergence of these structures -- the Kasparov module and semi-finite spectral triple -- in the form of three conjectures.

The importance  of having a semi-finite spectral triple with respect to an algebra generated by volume-preserving diffeomorphisms is that it guarantees invariant quantities. As long as we stick to the building blocks of the bi-module -- the Dirac type operator and the algebra -- then everything will remain invariant under these diffeomorphisms.\\

A key result presented in this paper and proven in \cite{AGnew} is that the separable part of the spectrum of $\mathbf{HD}(M) $ is contained in a space of measurable connections modulo gauge equivalence. As already mentioned this invites an interpretation in terms of Ashtekar connections and variables. This, in turn, brings us in contact with loop quantum gravity (LQG), which is also based on the Ashtekar connection and its holonomies. In this paper we devote a section to spell out the key similarities and differences between the two approaches. At a technical level, LQG is based on an algebra, which is constructed as a projective limit of algebras assigned to piece-wise analytic graphs. The spectrum of this algebra is different from the spectrum of $\mathbf{HD}(M) $ and its bulk consist of so-called generalized connections, non-separable objects which are absent in the spectrum of $\mathbf{HD}(M) $. The reason for this difference is that $\mathbf{HD}(M) $ captures information about the local measurable structure of the underlying manifold $M$ whereas the algebra in LQG does not. Interestingly, the generalized connections appear in our construction if we discretize $\mathbf{HD}(M) $. This means that the LQG spectrum arise from a discretized version of our construction. \\

This paper is the first of two papers concerned with the connection between the algebra of holonomy-diffeomorphisms and quantum gravity. In this paper we give a general exposition and in the second paper \cite{AGnew} we focus on mathematical details and in particular on the analysis of the spectrum of $\mathbf{HD}(M) $. \\

This paper is organized as follows: In section 2 we introduce key concepts from noncommutative geometry, in particular Kasparov modules and semi-finite spectral triples. Section 3 introduces canonical gravity formulated in terms of Ashtekar variables and their holonomies. The algebra of holonomy-diffeomorphisms is defined in section 4 where we give our two main theorems, which state that the separable part of spectrum of the algebra is contained in a space of measurable connections and which give restrictions to the non-separable part. In section 5 we introduce the conjugate variables to the holonomy-diffeomorphisms together with semi-classical states and conjecture the existence of a semi-finite spectral triple. Section 6 is concerned with the emergence of elements of fermionic quantum field theory and an almost commutative geometry in a semi-classical limit and section 7 gives a discussion on how the dynamics of quantum gravity might emerge from our constructions. Section 8 contains a comparison between the construction presented in this paper and that of loop quantum gravity. Section 9 contains a discussion of our results.

\section{Noncommutative geometry: Kasparov modules and spectral triples}
 \label{noncommutativesection}
In this section we introduce the key elements of noncommutative geometry which we shall need in the subsequent sections. For more details on noncommutative geometry we refer the reader to the two books \cite{ConnesBook} and \cite{ConnesMarcolliBook}. For background material on operator algebras we refer the reader to the books \cite{KR1,KR2} and \cite{Bratteli:1979tw,Bratteli:1996xq}.\\

In many situations in ordinary geometry, properties and quantities of a geometric space $X$ are  described dually via certain functions from $X$ to $\bbR$ or $\bbC$. Functions from $X$ to $\bbR$ or $\bbC$ come with a product, namely the pointwise product between two such functions. Due to the commutativity of $\bbR$ and $\bbC$ this product is commutative.        
One is, however, often led to consider cases where the product is noncommutative and where one would still like to apply methods and conceptual thinking of geometry. 
Noncommutative geometry is a framework that extents geometrical concepts and techniques to such cases.


\subsection{Noncommutative topology}

We start with topology. A possible noncommutative framework for topological spaces is the definition of  $C^*$-algebras. 
A $C^*$-algebra is an algebra $\cb$ over $\bbC$ with a norm $\| \cdot \|$, and an anti-linear involution $*$ such that $\|ab\|\leq \| a\|\| b\|$, $\|aa^* \|=\| a\|^2$ and $\cb$ is complete with respect to $\| \cdot \|$. 
A fundamental theorem due to Gelfand-Naimark-Segal states that $C^*$-algebras can be equally well defined  as norm closed $*$-invariant subalgebras of the algebra of bounded operators on some Hilbert space.    

The following theorem states that the concept of $C^*$-algebras is the perfect generalization of  locally compact Hausdorff-spaces.   
\begin{thm} \label{gn}
(\textit{Gelfand-Naimark})
$C_0(X)$ is a commutative $C^*$-algebra. 
Conversely, any commutative $C^*$-algebra has the form $C_0(X) $, where $X$ is a locally compact Hausdorff-space, and $C_0(X)$ denotes the algebra of continuous complex-valued functions on $X$ vanishing at infinity.

\end{thm} 

Given a commutative $C^*$-algebra the space $X$ for which $\cb=C_0(X)$ is given by
\begin{equation}X= \{ \gamma :\cb \to \bbC | \gamma \hbox{ nontrivial }C^*\hbox{-homomorphism} \}\nn  \end{equation}
equipped with the pointwise topology. The space $X$ is also called the spectrum of $\cb$.
 
This result shows that a topological space $X$ have an equivalent formulation in terms of the $C^*$-algebra of functions on $X$. The starting point of noncommutative geometry is to consider also noncommutative $C^*$-algebras as the noncommutative generalization of a topological space.

One then consider the spectrum of such algebras, which is defined as the set of all irreducible representations of $\cb$ on Hilbert spaces modulo unitary equivalence.

\subsection{Noncommutative Riemannian geometry }

The theorem by Gelfand and Naimark opens the door to noncommutative topology. To find a notion of noncommutative geometry we need the concept of a spectral triple. 

The crucial observation by Alain Connes is that given a compact manifold\footnote{Note that if the manifold is non-connected the geodesic distance can assume infinite values.} $M$ with a metric $g$, the geodesic distance $d_g$ of $g$, and thereby also $g$ itself, can be recovered by the formula 
\begin{equation} d_g(x,y)=\sup \{ |f(x)-f(y)| |f\in C^\infty (M)\hbox{ with } \| [ \not\hspace{-1mm} D,f]\| \leq 1   \}   , \label{connesdist}\nn
\end{equation}
where $ \not\hspace{-1,8mm} D$ is a Dirac type operator associated to $g$ acting in $L^2 (M,S)$, $S$ is some spinor bundle and $\|[ \not\hspace{-1.5mm} D,f]\|$ the operator norm of $[ \not\hspace{-1.5mm} D,f]$ as operator in $L^2 (M,S)$. Therefore to specify a metric, one can equally well specify the triple
\begin{equation} (C^\infty (M), L^2(M,S),  \not\hspace{-1mm} D). \nonumber\end{equation}

This observation leads to the definition:
\begin{definition} \label{spectrip}
A spectral triple $(\cb , \ch ,D)$ consists of a unital $*$-algebra   $\cb$ (not necessary commutative), a separable Hilbert space $\ch$, a unital $*$-representation 
\begin{equation}\varphi : \cb \to \cb (\ch) \nonumber\end{equation}
and a  self-adjoint operator $D$ (not necessary bounded) acting on $\ch$ satisfying
\begin{enumerate}  
\item $\frac{1}{1+D^2} \in \ck (\ch)$.
\item $[D,\varphi (b)]\in \cb(\ch)$ for all $b\in \cb  $. 
\end{enumerate}
\end{definition}
where $\ck(\ch)$ are the compact operators.
This definition is the replacement of metric spaces in the non-commutative setting. 

 Note that  the triple 
 $$(C^\infty (M), L^2(M,S),  \not\hspace{-1mm} D)$$ satisfies both conditions 1 and 2.  Property $1$ reflects the fact that the absolute values of the eigenvalues of $ \not\hspace{-1.8mm} D$ converges to infinity and that each eigenvalue only have finite degeneracy. Property $2$ reflects the fact that the functions in $C^\infty (M)$ are differentiable.

The definition \ref{spectrip} is insufficient as a definition of a non-commutative generalization of Riemannian geometry. In fact it can be shown that all compact metric spaces fit into \ref{spectrip}, see \cite{ChrisIvan}. Therefore to pinpoint a definition of a non-commutative Riemmanian  manifold one needs to add more axioms to those of a spectral triple. 

In \cite{ConnesRecon} it was shown that given a commutative spectral triple satisfying the extra axioms specified in \cite{ConnesRecon} it is automatically an oriented compact manifold. 
We will not give the details here but refer to \cite{ConnesRecon}. For a set of axioms for noncommutative oriented Riemmanian geometry see \cite{LRV},  for the original axioms of noncommutative spin manifolds see \cite{Connes:1996gi} and for a generalization to almost commutative geometries and a weakened orientability hypothesis, see \cite{cacic}. 

There is however one important aspect we want to mention here. In noncommutative spin geometry there is an extra ingredient, the real structure $J$, which plays an important role. For a four dimensional spin manifold $J$ is the charge conjugation operator. In general $J$ is required to be an anti-linear operator on $\ch$ with the property that $Ja^*J^{-1}$ gives a right action of $A$ on $\ch$, and satisfying some additional axioms.

\subsection{The standard model}   
Perhaps the most intriguing outcome of noncommutative geometry is the natural incorporation of the standard model of particle physics coupled to gravity into the framework, and in particular the severe restrictions this puts on other possible models in high energy physics. Since the details of this are very subtle and elaborate,  we will here merely provide a sketch and refer the reader to \cite{Chamseddine:2006ep,Chamseddine:2012sw} (for a recent review aimed at a physics audience see also \cite{Dungen:2012ky}).

The basics of the construction is to combine the commutative spectral triple      
\begin{equation}(C^\infty (M), L^2(M,S),  \not\hspace{-1mm}D ),\nonumber\end{equation} where $M$ is a 4-dimensional manifold, with a finite dimensional triple 
$$(\ca_F , \ch_F, D_F)\;,$$ 
where $\ca_F$ is a matrix algebra, by tensoring them, i.e.
\begin{equation}(C^\infty (M)\otimes \ca_F , L^2(M,S)\otimes \ch_F ,  \not\hspace{-1mm} D\otimes 1 +\gamma_5\otimes D_F) . \nonumber\end{equation}
Of course the exact structure of $(\ca_F , \ch_F, D_F)$ is to a large degree dictated by the structure of the standard model. The Hilbert space $\ch_F$ labels the fermionic content of the standard model. Elements $\psi \in  L^2(M,S)\otimes \ch_F$ describe the fermionic fields.

Given this triple the noncommutative differential forms, which for a spectral triple $(\cb,\ch,D)$ are generally of the form
\begin{equation}
 a_i[ \not\hspace{-1mm} D,b_i] , \quad a_i,b_i \in \cb\;,\nonumber
 \end{equation}
generate the gauge sector of the standard model and the action of the standard model minimally coupled to the Euclidean background given by $\not\hspace{-1mm} D$ is given by
\begin{equation}\cl (A,\psi ) =\Tr \phi\left( \frac{D_A^2}{\Lambda^2} \right) +\langle J\psi | D_A\psi \rangle ,\nonumber\end{equation}
where $\phi$ is a suitable function, $\Lambda$ is a cutoff and $D_A$ is a Dirac operator obtained via inner automorphisms of $\cb$ and $A$ is a one-form.

\subsection{Unbounded Kasparov modules and semi-finite spectral triples}

The definition of a spectral triple may in some cases be found to be inadequate due to the presence of a symmetry group, which entails a degenerate spectrum of the Dirac type operator. To deal with such cases we need a machinery where everything works up to an action of that group. The first component hereof is the notion of a Hilbert $C^*$-module \cite{Blackadar}
\begin{definition}
Let $B$ be a $C^*$-algebra. A pre-Hilbert $B$-module is a complex linear space $E$ equipped with a compatible right $B$-module structure and a map 
$$
\langle \cdot,\cdot\rangle: E \times E \rightarrow B
$$
satisfying:
$$
\langle x, \a y + \b z \rangle = \a \langle x,y\rangle + \b \langle x,z\rangle  \;,
\quad 
\langle x,yb\rangle = \langle x,y\rangle b  \;,
\quad 
\langle x,y \rangle = \langle y,x \rangle^* \;,
$$
where $x,y,z\in E$, $\a,\b\in\mathbb{C}$ and $b\in B$.
Define a norm on $E$ by
$$
\vert\vert x \vert\vert = \vert\vert  \langle x,x\rangle  \vert\vert^{\frac{1}{2}} \;.
$$
The norm completion of $E$ is then called a Hilbert $B$-module.
\end{definition}
In this definition $B$ can be thought of as an algebra generated by a redundant symmetry group. Thus, a Hilbert $C^*$-module is essentially a Hilbert space "up to an algebra $B$ generated by a redundant symmetry group". 

Next we need a way to deal with an unbounded, self-adjoint operator -- to be a Dirac type operator -- with a spectrum that is degenerate with respect to $B$. The right framework for this is that of an unbounded Kasparov module. The following definition is taken from \cite{Blackadar}
\begin{definition}
Let $A$ and $B$ be graded $C^*$-algebras. An unbounded Kasparov module for $(A,B)$ is a triple $(E,\phi,D)$ where $E$ is a Hilbert $B$-module, $\varphi: A\rightarrow {\bf B}(E)$ is a graded $^*$-homomorphisms and $D$ is a self-adjoint regular operator on $E$, homogeneous of degree 1, such that
\begin{itemize}
\item[(i)]
$(1+D^2)^{-1} \varphi(a)$ extends to an element of ${\bf K}(E)$ for all $a\in A$.
\item[(ii)]
The set of $a\in A$ such that $[D,\varphi(a)]$ is densely defined and extends to an element of ${\bf B}(E)$, is dense in $A$.
\end{itemize}
\end{definition}
Thus both the operator $D$ and the algebra $A$ commute with $B$ corresponding to the degeneracy. Note that in this paper we shall only deal with cases where $A$ and $B$ have trivial gradings.

The next step is to turn the Hilbert module into a Hilbert space and $D$ into a self-adjoint, unbounded operator hereon, viz. a Dirac type operator. To do this one needs to have a normalized trace on $B$, which promotes the map $\langle\cdot,\cdot\rangle$ to a Hilbert space inner product. Hereby the eigen-projections of $D$, which takes value in the compact operators over $B$, will be normalized by the trace.  
The key concept to achieve this is that of a semi-finite spectral triple \cite{Carey}: 
\begin{definition}
Let $\cn$ be a semi-finite von Neumann algebra with a semi-finite trace $\t$. Let $\ck_\t$ be the $\t$-compact operators. A semi-finite spectral triple $(A,H,D)$ is a *-sub-algebra $A$ of $\cn$, a representation of $\cn$ on the Hilbert space $H$ and an unbounded densely defined self-adjoint operator $D$ on $H$ affiliated with $\cn$ satisfying:
\begin{enumerate}
\item
$a(\l-D)^{-1}\in \ck_\t$ $ \forall a\in A$ and $\l\in\mathbb{R}$
\item
$[a,D]$ is densely defined and extends to a bounded operator. 
\end{enumerate}
\end{definition}
A von Neumann algebra $\cn$ is an algebra that can be written as a double commutant of some other algebra and a semi-finite trace on $\cn$ is, essentially, a trace which is finite on a dense sub-algebra of $\cn$.

Given a Kasparov $A$-$B$-module and a trace over $B$ one has is in fact automatically a semi-finite spectral triple. The trick is extend the trace on $B$ to a trace on a von Neumann algebra $\cn$ that includes both $A$ and $B$. We refer to \cite{RenniePask} for the details on how this is accomplished.

Once a semi-finite spectral triple is constructed one has a machinery which produces well defined quantities despite the degeneracy generated by the algebra $B$.

\section{Connection formalism of gravity}
\label{confor}

Let us start by recalling the formulation of canonical gravity in terms of connection variables (for details see \cite{AL1}). 
First assume that space-time $\cm$ is globally hyperbolic. Then $\cm$ can be foliated as 
$$\cm=M \times \mathbb{R}, $$   
where $M$ is a three dimensional hyper surface. 
The fields in which we will describe gravity are the Ashtekar variables\footnote{The original Ashtekar connection is a complexified $SU(2)$ connection. Within LQG it is, however, custom to work with a real $SU(2)$ connection due to restrictions arising from the construction of the kinematical Hilbert space. This choice of gauge group alters the constraint algebra (\ref{constraints}). In this paper we shall likewise restrict ourselves to a real connection although the restrictions found in LQG do not arise here. We shall comment on this issue in section \ref{Discussion}.} \cite{Ashtekar:1986yd,Ashtekar:1987gu} given by
\begin{itemize}
\item a $SU(2)$-connection $\nabla$ in the trivial bundle over $M$. We denote by $A_i^a$ the local 1-form of $\nabla$, where $a$ is the $\mathfrak{su}(2)$-index.
\item a $\mathfrak{su}(2)$-valued vector density on $M$. We will adopt the notation $E_a^i$.  
\end{itemize} 
On the space of field configurations, which we denote $\mathcal{P}$, there is a Poisson bracket expressed in local coordinates by
$$\{ A_i^a (x), E_b^j(y) \} =\delta_i^j \delta_b^a \delta (x,y), $$
where $\delta (x,y)$ is the delta function on $M$. The rest of the brackets are zero.
These fields are subjected to constraints given by
\begin{eqnarray}
 \epsilon_c^{ab} E^i_a E^j_b F_{ij}^c&=&0\;\nonumber\\
   E^j_a F^a_{ij}&=&0  \;\nn \\
  (\partial_i E^i_a+\epsilon_{ab}^cA_i^b E^i_c)&=&0\;.
\label{constraints}
\end{eqnarray}
Here $F$ is the field strength tensor of the connection $A$. The first constraint is the Hamilton constraint, the second is the diffeomorphism constraint and the third is the Gauss constraint.
These field configurations together with the constraints constitute an equivalent formulation of General Relativity without matter.

\subsection{Holonomy and flux variables}
The formulation of gravity in terms of connection variables permit a reformulation of the Poisson bracket in terms of holonomies and fluxes. For a given path $p$ in $M$ the holonomy function is simply the parallel transport along the the path, i.e.
$$
\mathcal{P}\ni (A,E) \to h_p(A)\in G\;,
$$
where $G$ is the gauge group. Given an oriented surface $S$ in $M$ the associated flux function is given by
$$
\mathcal{P}\ni (A,E)\to E^S_a := \int_S \epsilon_{ijk} E_a^idx^jdx^k \;.
$$

Let $p$ be a path and $S$ be an oriented surface in $M$ and assume $p$ ends in $S$ and has exactly one intersection point with $S$. The Poisson bracket in this case becomes 
\begin{equation}
\{ h_p, E^S_a \}(A,E)=\pm \frac{1}{2} h_p(A)\sigma_a 
\label{poisson1}
\end{equation} 
where $\sigma_a$ is the Pauli matrix with index $a$. The sign in (\ref{poisson1}) is negative if the orientation of $p$ and $S$ is the same as the orientation of $M$, and positive if not. If $p$ instead starts on $S$ one gets the Poisson bracket
\begin{equation}
\{ h_p, E^S_a \}(A,E)=\pm \frac{1}{2}\sigma_a h_p(A) \label{poisson2}
\end{equation}
but now with the reverse sign convention.
If $p$ is contained in $S$,  if $p$ fails to intersect $S$ or if $p$ end at $S$ with a velocity vector tangent to $S$, the Poisson bracket is zero.

\section{The holonomy-diffeomorphism algebra}

We are now ready to introduce the holonomy-diffeomorphism algebra, which will be the cornerstone in the construction presented in this paper. In this section we give a definition of the algebra and state the two main theorems on its spectrum, and in the next section we construct a semi-finite spectral triple over the algebra.


A more complete analysis of the holonomy-diffeomorphism algebra and its spectrum is given in \cite{AGnew}, where we also give a more general definition of the algebra.\\ 

Let $S$ be a spin bundle over $M$. We assume that $S$ is equipped with a fibre wise metric. This metric ensures that we have a Hilbert space $L^2 (M , \Omega^{\frac12} \otimes S)$, where $\Omega^{\frac12}$ denotes the bundle of half densities on $M$. Given a diffeomorphism $\phi: M\to M$ this acts unitarily on  $L^2 (M , \Omega^{\frac12} )$ through
$$ \phi (\xi)(m)= \phi^*(\xi (\phi (m) )  , $$
where 
$$\phi^* :\Omega^{\frac12} (\phi (m)) \to \Omega^{\frac12} (m)  $$
denotes the pullback.

Let $X$ be a vectorfield on $M$, which can be exponentiated, and let $\nabla$ be a connection in $S$.  Denote by $t\to \exp_t(X)$ the corresponding flow. Given $m\in M$ let $\gamma$ be the curve  
$$\gamma (t)=\exp_{1-t} (X) (\exp_1 (X)(m) )$$
running from $\exp_1 (X)(m)$ to $m$. We define the operator 
$$e^X_\nabla :L^2 (M , \Omega^{\frac12} \otimes S) \to L^2 (M , \Omega^{\frac12} \otimes S)$$
in the following way: let us consider an element $\xi \in L^2(M,\Omega^{\frac12} \otimes S)$, and let us assume that we write it locally over $(\exp_1(x)(m))$ as  $$f \otimes \omega \otimes s\;,$$ where $f$ is a function, $\omega$ an element in $\Omega^{\frac12}$ and $s$ an element in $S$. The "value" of $(e^X_\nabla)(\xi)$ in the point  $m$ is given as
$$
(f((\exp_1(X))(m))) \otimes (\exp_1^*(\omega)) \otimes ((h_\gamma (\nabla) )s)
$$
where $h_\gamma(\nabla)$ is the holonomy of $\nabla$ along $\gamma$.
If the connection $\nabla$ is unitary with respect to the metric on $S$, then  $e^X_\nabla$ is a unitary operator. 

If we are given a system of unitary connections $\ca$ we define an operator valued function over $\ca$ via
$$\ca \ni \nabla \to e^X_\nabla    ,$$
and denote this by $e^X$. Denote by $$\cf (\ca , \cb (L^2(M,\Omega^{\frac12}\otimes S)) )$$ the bounded operator valued functions over $\ca$. This form a $C^*$-algebra with the norm
$$\| \Psi \| =  \sup_{\nabla \in \ca} \{\|  \Psi (\nabla )\| \}. $$ 
  
For a function $f\in C^\infty_c (M)$ we get another operator valued function $fe^X$ on $\ca$.

\begin{definition}
Let 
$$C =   \hbox{span} \{ fe^X |f\in C^\infty_c(M), \ X \hbox{ exponentiable vectorfield }\}  . $$
The holonomi-diffeomorphism algebra $\mathbf{HD}(M,S,\ca) $ is defined to be the $C^*$-subalgebra of  $\cf (\ca , \cb (L^2(M,\Omega^{\frac12}\otimes S)) )$ generated by $C$.
\end{definition}


For simplicity we shall also denote the holonomy-diffeomorphism algebra by $\mathbf{HD}(M) $.

Notice that $\mathbf{HD}(M)$ is the completion of
the semi-direct product
$$
C^\infty_c (M)  \rtimes  \cF  / I\;,
$$
in the norms described above, where $\cf$ is the group generated by flow operators $e^X$, $I$ is an ideal given by certain reparametrizations of flows (see \cite{AGnew} for details) and where $C^\infty_c (M) $ is the algebra of smooth functions with compact support. The semi-direct product comes with the multiplication relation
$$f_1F_1 f_2 F_2=f_1 F_1 (f_2) F_1 F_2 \;,  $$
where $F_1,F_2\in\cf$.

\subsection{The spectrum of $\mathbf{HD}(M) $}

Our first concern is to determine the spectrum of the algebra $\mathbf{HD}(M) $, which is defined as 
 the irreducible representations of $\mathbf{HD}(M) $ modulo unitary equivalence. We have two main results concerning the spectrum of $\mathbf{HD}(M) $. Before we give these we need to introduce the concepts of a measurable connection and of a generalized connection.

\begin{definition}
A measurable $U(n)$-connection, $n=1,\ldots , \infty$, is a map $\nabla$ from $\cF$ to the group of measurable maps from $M$ to $U(n)$ satisfying
\begin{enumerate}
\item $\nabla (1)= 1$.
\item $\nabla (F_1 \circ F_2)(m)=\nabla (F_1) (m) \circ \nabla (F_2)(F_1^{-1}(m))$
\item If $F_1$ and $F_2$ are the same up to local reparametrization over some set $U\subset M$, then 
$$ \nabla ( F_1)_U= \nabla (F_2)_U  . $$ 
\end{enumerate} 
\end{definition}

Let $l$ be a piece-wise analytic path in $M$. We identify $l\cdot l^{-1}$ with the trivial path starting and ending at the start point of $l$. Furthermore we identify two paths that differ by a reparameterization. 
\begin{definition}
Let $G$ be a connected Lie-group. A generalized connection is an assignment $\nabla(l)\in G$ to each piece-wise analytic path $l$, such that
$$
\nabla(l_1\cdot l_2) = \nabla(l_1)\nabla(l_2)\;.
$$
\end{definition}

We can now state our two main results on the spectrum of $\mathbf{HD}(M) $:
\begin{thm}
\label{TH1}
Any separable, irreducible representation of $\mathbf{HD}(M) $ is unitarily equivalent to a representation of the form $\varphi_\nabla$, where $\nabla$ is a measurable $U(2)$-connection\footnote{Here we disregard the special case where the representation decomposes into two $U(1)$ measurable connections.}.
\end{thm}

{\it Proof:} see \cite{AGnew}.\\

\begin{thm}
\label{TH2}
A generalized connection together with the counting measure on $M$ does not render a representation of $\mathbf{HD}(M) $.
\end{thm}

{\it Proof:} see \cite{AGnew}.\\

The proof of theorem \ref{TH1} is based on the fact that a separable representation of an algebra, which has $C^\infty(M)$ as a subalgebra, necessarily has a Hilbert space that includes $L^2(M)$, see \cite{Kadison}. Once this is established one uses the action of the diffeomorphisms to control the spectral multiplicity of the representation, which means that a representation which is $n$-dimensional in a point in $M$ will  be so all over $M$.

Theorem \ref{TH1} holds in more general settings -- manifolds of arbitrary dimensions and arbitrary vector bundles. The theorem is, however, particularly interesting in the case where $M$ is a three-dimensional manifold and $S$ is a two-spinor  bundle over $M$ with $SU(2)$ connections, since in this case one can interpret the spectrum as the completion of a configuration space of Ashtekar connections. 
For the remainder of this paper we shall define $\mathbf{HD}(M) $ as the algebra generated by flows acting on a two-spinor bundle with $SU(2)$ connections.

It is an open question what the non-separable part of the spectrum of $\mathbf{HD}(M) $ contains. The fact that we are for now unable to prove that the entire spectrum is given by measurable connections may indicate that we need to change the definition of $\mathbf{HD}(M) $. In particular, we think that the topology of $\mathbf{HD}(M) $, which is the $C^*$-topology, is not the right one for our purpose. For further discussion of this point see the end of section \ref{sec51} and section \ref{Discussion}.

Theorem \ref{TH2} basically states that the bulk of the spectrum found in loop quantum gravity, which is given by generalized connections with support on finite graphs, is excluded from the spectrum of $\mathbf{HD}(M) $.

Note that although we may start with a non-trivial bundle it will, in the spectrum of $\mathbf{HD}(M) $, always be trivial. This is due to the measurable nature of the spectrum, which will do away with any topological obstruction.


Note also that there is an open question as to how we get $SU(2)$ connections instead of $U(2)$ connections. Of course, we can simply put them in by hand -- which is what we do in this paper -- but a more natural solution would be to introduce a real structure and a conjugate action of the algebra, which would kill the $U(1)$ factor.

Also, it might be interesting to consider  connections with a non-compact structure group\footnote{The original Ashtekar connection is a complexified $SU(2)$ connection and therefore has a non-compact structure group. See section \ref{Discussion} for a brief discussion.}. In this case one would need to consider infinite-dimensional unitary representations.

\section{A spectral triple over $\mathbf{HD}(M) $ }

The algebra $\mathbf{HD}(M) $ involves one part of the canonical variables introduced in section \ref{confor} -- the holonomies. In this section we introduce their canonical conjugates, the densitized triad fields. 
To do this we need to introduce an alternative construction of $\mathbf{HD}(M) $, which is based on a coordinate system in $M$. This will also allow us to construct semi-classical states on $\mathbf{HD}(M) $ and to study their GNS representations.

The introduction of a coordinate system in $M$ clearly raises the question whether the construction is background independent. This question turns out to reveal the raison d'\^etre for organizing the quantized triad fields in a Dirac type operator: it turns out that this is an operator invariant under -- at least -- volume-preserving diffeomorphisms.

\begin{figure}[t]
\begin{center}
\resizebox{!}{2.5cm}{
 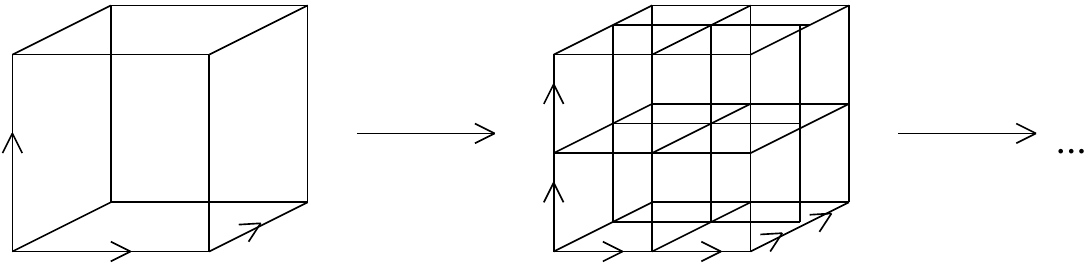}
\end{center}
\caption{\it One subdivision of a cubic lattice.}
 \label{ronnie}
\end{figure}

\subsection{Reformulating $\mathbf{HD}(M) $ in terms of lattices}
\label{sec51}

Let $\OO=\{\G_n\}$ be an infinite sequence of 3-dimensional, nested, cubic lattices in $M$, see figure \ref{ronnie}. Given a lattice $\G_n$ denote by $\{v_i\}$ and $\{l_j\}$ its vertices and edges. 
Consider now sequences of adjacent edges $\{l_{i_1}, l_{i_2},\ldots l_{i_n}\}$ in $\G_n$.  Two sequences in $\G_n$ that differ by a trivial backtracking are said to be equivalent
$$
\{\ldots l_{i_k},l_{i_{k+1}}\ldots l_{i_{k+j}} ,l^*_{i_{k+j}} \ldots l^*_{i_{k+1}},l_{i_{k+j+1}} \ldots \}\sim\{\ldots l_{i_k},l_{i_{k+j+1}} \ldots \}
$$
where $l_i^*$ is the reverse of $l_i$. This is an equivalence relation and we call an equivalence class for a path.

\begin{figure}[t]
\begin{center}
\resizebox{!}{4cm}{
 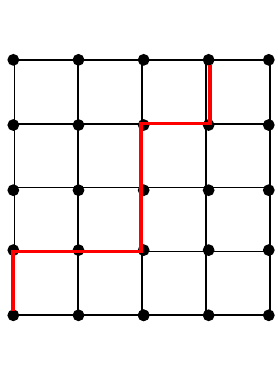}
\end{center}
\caption{\it A path in $\G_n$ connecting $v_0$ and $v_1$. $\mathbbm{F}_n$ is a family of such paths.}
 \label{ronnie2}
\end{figure}

Denote by $\mathbbm{F}_n$ a finite family of oriented paths running in $\G_n$, see figure \ref{ronnie2}. The set of such families is a groupoid, where the product is given by composition of paths. Given a $SU(2)$ connection $\nabla$ in a principal bundle over $M$ we shall define an action of $\mathbbm{F}_n$ as an operator in $L^2(M,S)$ in the following manner: first, we subdivide $M$ into cubes $\{c_i\}$, where a cube $c_i$ is assigned to each vertex $v_i\in\G_n$. Let us assume that $\mathbbm{F}_n$ includes a path $p$ which connects two vertices $v_i$ and $v_j$. Then $\mathds{F}_n$ acts on a spinor $\psi\in L^2(M,S)$ by shifting each value of $\psi$ in the cube $c_i$ to the same relative location in the cube $c_j$ while multiplying it with the holonomy transform $h_p(\nabla)$ of the connection along $p$. We denote this representation by $\varphi(\mathbbm{F}_n)$.

Next, denote by $\mathbbm{fF}_n$ a finite family of oriented path where we also assign to each path a weight. $\mathbbm{fF}_n$ has the same representation, denoted $\varphi(\mathbbm{fF}_n)$, in $L^2(M,S)$ as $\mathbbm{F}_n$ except that each parallel transport is also multiplied with the weight.

 $\mathbbm{F}_n$ should be understood as an approximation of an element $F$ in the flow group $\cf$ and $\mathbbm{fF}_n$ as an approximation of an element $fF$ in  $ \mathbf{HD}(M) $, where the weight plays the role of a discretization of a function $f$.

Denote by $\{\mathbbm{fF}_n\}$  an infinite sequence of families $\mathbbm{fF}_n$ of paths associated to each $\G_n$ in $\OO$.
We say that a representation $\varphi$ of $\{\mathbbm{fF}_n\}$ converges to $fF\in \mathbf{HD}(M) $ if
\begin{equation}
\lim_{n\rightarrow\infty}\vert\vert (\varphi(\mathbbm{fF}_n) - \varphi(fF)\xi \vert\vert = 0\;,
\label{xtxt}
\end{equation}
for all $\xi\in L^2(M,S)$ and for all smooth connections,
where $\varphi(fF)$ denotes the representation of $fF$ in $L^2(M,S)$. Keep in mind that the representation $\varphi$ depends on a chosen $SU(2)$ connection and that (\ref{xtxt}) must hold for any such choice.
There is a natural equivalence relation between the sequence $\{\mathbbm{fF}_n\}$ and $\{\mathbbm{fF}'_n\}$ given by
\begin{equation}
\{\mathbbm{fF}_n\}\sim \{\mathbbm{fF}_n'\}\quad iff \quad \lim_{n\rightarrow\infty}\vert\vert ({\varphi(\mathbbm{fF}_n)} - \varphi(\mathbbm{fF}_n'))\xi \vert\vert = 0
\label{xtxtx}
\end{equation}
for all $\xi\in L^2(M,S)$ and all smooth connections. We will, when we discuss infinite sequences $\{\mathbbm{fF}_n\}$, implicit assume that we are considering such equivalence classes.
If a representation $\varphi$ of $\{\mathbbm{fF}_n\}$ converges to $fF$ we 
shall say that  $\varphi(\mathbbm{fF}_n)$ approximates $fF$.

Before we continue let us introduce a second representation of $\{\mathbbm{fF}_n\}$ in $L^2(M,S)$, which corresponds to diffeomorphisms on $M$ -- we let $\mathbf{D}(M) $ denote the diffeomorphisms on $M$.  Let $\varphi^o(\mathbbm{fF}_n)$ be an operator in $L^2(M,S)$, which is identical to $\varphi(\mathbbm{fF}_n)$ except that we do not multiply with the holonomy transform of the connection. Thus, this representation is independent of the path and of the connection. Given a local  diffeomorphism $\phi$ in $\mathbf{D}(M) $, which coincide with an element $fF$ in $\mathbf{HD}(M) $ when the latter acts on scalar functions, we say that $\varphi^o(\mathbbm{fF}_n)$ approximates $\phi$ iff
\begin{equation}
\lim_{n\rightarrow\infty}\vert\vert \varphi^o(\mathbbm{fF}_n) (\xi)  - \phi(\xi)  \vert\vert = 0\;,
\label{xtxtxt}
\end{equation}
for all $\xi\in L^2(M,S)$ and for all smooth connections,
and again we identify equivalent sequences.
We note that if a representation $\varphi$ of $\{\mathbbm{fF}_n\}$ converges to an element in $\mathbf{HD}(M) $ then the corresponding representation $\varphi^o$ will approximate the corresponding diffeomorphism in $\mathbf{D}(M) $. Of course, the diffeomorphisms are not path dependent.

Finally, let us also introduce operators, which correspond to diffeomorphisms that preserve the volume of $M$ with respect to the background metric given by the coordinate system $\OO$. We denote these volume-preserving diffeomorphisms by $ \mathbf{D}_{vol}(M) $. These are constructed like elements in $ \mathbf{D}(M) $ except that we only permit flows $\mathbbm{F}_n$ that are invertible.

We denote by $\mathbf{HD}_n(M) $, $\mathbf{D}_n(M) $ and $\mathbf{D}_{vol,n}(M) $ the $^*$-algebras which approximate $\mathbf{HD}(M) $, $\mathbf{D}(M) $ and $\mathbf{D}_{vol}(M) $ at the level of $\G_n$.\\

It is important to realize the implication of working with equivalence classes of sequences.  The difference between two equivalent sequences correspond to quantities, which have zero measure with respect to $L^2(M)$. It is exactly the intent of our constructions to remove any dependency on such quantities\footnote{The existence of a continuum limit in which zero-measure quantities vanish was first conjectured in \cite{Aastrup:2009ra}.}. \\

Notice that we have formulated the convergence conditions (\ref{xtxt}),  (\ref{xtxtx}) and  (\ref{xtxtxt}) with respect to the strong topology. This means that we are in fact not approximating the holonomy diffeomorphism algebra $\mathbf{HD}(M) $ but rather elements hereof embedded in a larger von Neumann algebra. It is necessary to use the strong topology since it has the sensitivity to the topology of $M$ which a lattice approximation requires. The $C^*$-topology is only partly sensitive to the topology of $M$.

This issue seems to suggest that we have chosen the wrong topology for $\mathbf{HD}(M) $ and that we should define the holonomy-diffeomorphism algebra with the strong topology. This would entail a von Neumann algebra setting. For now we will merely point out this discrepancy in our construction and leave it for future publications to determine exactly which algebra topology -- and therewith what algebra --  is most suitable for our purpose. See also section \ref{Discussion} for further discussions of this issue.

\subsection{Intermediate spaces of connections and Hilbert spaces}

Let again $\{\mathbbm{fF}_n\}$ be a sequence which converges to $fF\in \mathbf{HD}(M) $. Assign to each lattice $\G_n$ the space
$$
\ca_n = G^{\# \G_n}
$$
where $G$ is a locally compact Lie group and where $\# \G_n$ is the number of edges in $\G_n$. Thus, we assign one copy of $G$ to each edge in $\G_n$. In the following we shall assume $G$ to equal $SU(2)$. 
Denote by $\nabla$ a map which assigns an element of $G$ to each edge in $\G_n$
$$
\nabla(l_i)= g_i\in G
$$
and notice that $\ca_n$ is the space of all such maps. We extend the action of $\nabla$ to paths by
$$
\nabla(p) := \nabla(l_{i_1})\nabla( l_{i_2})\ldots\nabla( l_{i_n})
$$
and notice that this respects the equivalence relation given by trivial backtracking since 
$$
\nabla(l_i^*)=\nabla(l_i)^*\;.
$$
The $^*$-operation is an involution.
Next, consider the Hilbert space
\begin{equation}
H'_n = L^2(\ca_n,M_2(\mathbb{C}))\times M_{\vert {\bf v}\vert}(\mathbb{C})
\label{H1}
\end{equation}
where $L^2$ is with respect to the Haar measure on $G^{\# \G_n}$ and where $\vert {\bf v}\vert$ is the number of vertices in $\G_n$. There is a natural action of a path $p$, which runs from $v_i$ to $v_j$, on $H'$ given by
$$
h_p\cdot \x (\nabla) (v_i,v_k) = \nabla(p)\cdot \x (\nabla) (v_j,v_k)\;, \quad \x\in L^2(G^{\# \G_n},M_2(\mathbb{C}))\;,
$$
where '$\cdot$' is matrix multiplication in the $M_2(\mathbb{C})$ factor and where we assumed that we have chosen a two-by-two representation of $G$. 
We can now represent $\mathbbm{fF}_n$ as an operator in $H'$ by organizing it in a $\vert {\bf v}\vert \times  \vert {\bf v}\vert$-matrix, which acts on the $M_{\vert {\bf v}\vert}(\mathbb{C})$ factor of $H'_n$. We shall also denote this representation by $\varphi(\mathbbm{fF}_n)$.

Let $\{\mathbbm{fF}_n\}$ converge to $fF\in \mathbf{HD}(M) $ and let $\x$ be a state on $\mathbf{HD}(M) $. Let $\{\x_n\}$ be a sequence of states with $\x_n\in H'_n$. We will say that $\x_n$ approximates $\x$ if
$$
\langle \mathbbm{fF}_n \x_n\vert \x_n\rangle \rightarrow \x(fF)
$$
for all elements in the equivalence class $\{\mathbbm{fF}_n\}$. We say that two sequences $\{\x_n\}$ and $\{\x_n'\}$ are equivalent if they both approximate the same state $\x$. We will, when discussing states and their approximations always assume that we are dealing with such equivalence classes. 

\subsection{Vectorfields and a Dirac operator on $\G_n$}
\label{hulk}

Consider again a finite lattice $\G_n$ and the corresponding space $\ca_n=G^{\# \G_n}$. We now choose a left and right invariant metric $\langle \cdot,\cdot\rangle$ on G. We will consider the corresponding metric on $T^*G$. 
Equip $T^*\ca_n$ with the product metric. 
$Cl(T^*\ca_{n})$ is the Clifford bundle with respect to this metric.
Denote then by
$$
H_n = L^2(\ca_n,Cl(T^*\ca_{n})\otimes M_2(\mathbb{C}))\times M_{ \vert {\bf v} \vert}(\mathbb{C})
$$
a modification of the Hilbert space (\ref{H1}).

Let $\mathfrak{g}_i$ be the Lie algebra of the $i$'th copy of $G$ and choose an orthonormal basis $\{e_i^a\}$ for $\mathfrak{g}_i$ with respect to $\langle \cdot,\cdot\rangle$. Here the index $i$ labels the different copies of $G$ while $a$ is an index for the Lie algebra of $G$.  We also denote by $\{{\bf e}_i^a\}$ the corresponding left translated vectorfields and by $\cl_{{\bf e}_i^a}$ the derivation with respect to the trivialization given by $\{e_i^a\}$. 

The vector field $\cl_{{\bf e}_i^a}$ corresponds to a quantized flux variable $E^S_a(x)$ where '$x$' is the endpoint of the $i$'th edge and $S$ is a surface that intersects $l_i$ at $x$ (this surface play no role in the following). To see this one simply computes the commutator between $\cl_{{\bf e}_i^a}$ and $h_{p_i}$, where $p_i$ is the path which consist of the edge $l_i$, and compare it to the Poisson bracket (\ref{poisson1}) (see \cite{AGNP1} and \cite{Aastrup:2009ra} for details). 

In order to obtain a representation of the quantized flux variables in the setup described here we could introduce infinite sequences of vector fields $\cl_{{\bf e}_i^a}$ where each vector field is associated to a finite graph in a manner such that the involved edges would converge onto a point in $M$. Such sequences would then interact with the flow algebra $\mathbf{HD}(M) $ and the combination of the two capture the kinematics of quantum gravity.  Such sequences would play the role as quantized flux operators.

In the following we will, however, adopt a slightly different approach. We will instead construct a candidate for a Dirac type operator from the vector fields $\cl_{{\bf e}_i^a}$. This operator will be an infinite sequence of intermediate Dirac type operators, each associated to a finite graph, and it will represent a kind of integrated operator over all possible flux operators. The reason for doing this -- as will be explained in the following -- is that such an operator will be invariant under a large class of diffeomorphisms. 

Consider again a graph $\G_n$. The Dirac type operator is written
\begin{equation}
D_n = \sum_{a,i} \a_n e_i^a \cdot \cl_{{\bf e}_i^a} \otimes \mathds{1}_{\vert {\bf v}  \vert }  \;,
\label{DIRAC}
\end{equation}
where $\a_n$ is an arbitrary real positive constant, where '$\cdot$' denotes Clifford multiplication and where $\mathds{1}_{\vert {\bf v}\vert}$ is a $\vert {\bf v}  \vert \times \vert {\bf v}  \vert $ identity matrix.  
Then the triple
\begin{equation}
(\mathbf{HD}_n(M) ,H_n,D_n)\;,
\label{diracie}
\end{equation}
form a spectral triple associated to the finite graph $\G_n$.

 Notice that the operator (\ref{DIRAC}) commutes with the action of the $^*$-algebra $\mathbf{D}_{vol,n}(M)$ of volume-preserving diffeomorphisms  on the graph $\G_n$. 
If we had let $\a_n$ depend on the index $i$ as well, which means that each copy of $G$ would have a different weight, then $D_n$ would not commute with elements in $\mathbf{D}_{vol,n}(M) $. 

At a first sight one would also say that $D_n$ commutes with all elements in  $\mathbf{D}_{n}(M) $ since these only act on the $\vert {\bf v}  \vert \times \vert {\bf v}  \vert $ matrix part of $H_n$. There is, however, an additional action by diffeomorphisms on the configuration space $\ca$ of smooth connections (alternatively: on the spectrum of $\mathbf{HD}(M)$) and therefore there is also an induced action of both $\mathbf{D}_{vol,n}(M) $ and $\mathbf{D}_{n}(M) $ in $L^2(\ca_n)$. Here it is only the volume preserving diffeomorphisms that conserve the number of copies of $G$ in $\ca_n$. Therefore the Dirac type operator $D_n$ will -- apparently -- only commute with this type of diffeomorphisms.

Notice also that the operator (\ref{DIRAC}) is essentially identical\footnote{The Dirac type operator, which we constructed in \cite{AGN3}, had to be compatible with a system of Hilbert space embeddings related to subdivision of graphs. This requirement was solved by having the sum over copies of $G$ in $\ca_n$ run with respect to a certain coordinate system hereon. Also, the scaling factors $\a_n$ were differently arranged and the Hilbert space did not involve points in $M$ and thus there was no $\mathds{1}_{\vert {\bf v}  \vert } $ in the Dirac type operator.} to the operator which we first introduced in \cite{AGN3}.\\

It is at the present level of analysis far from obvious that one can take a sensible continuum limit $n\rightarrow\infty$ of a sequence of triples in (\ref{diracie}). For instance, it is not clear how one can define a limit of the Hilbert spaces\footnote{In \cite{AGN1}-\cite{AGN3} we used projective and inductive limits to define a spectral triple which involved all graphs in $\OO$. This strategy will, however, not work in the setup considered here since elements in $\mathbf{HD}(M)$ are represented as equivalence classes of infinite sequences. We need a construction where finite approximations have no importance and only the continuum limit matters. } $H_n$. Indeed, instead of trying to do so we shall adopt a different strategy. In the next section we introduce coherent states on the flow algebra $\mathbf{HD}(M) $, and it is the GNS construction of these states which shall provide us with a Hilbert space structure.

\subsection{Semi-classical states on $\mathbf{HD}_n(M) $ and $\mathbf{HD}(M) $}

We start by recalling results for coherent states on various copies of $SU(2)$. This construction uses results of Hall \cite{H1,H2} and is inspired by the articles \cite{BT1}-\cite{BT2}.

First pick a point $(A_n^a,E^m_b)$ in the phase space $\cp$ of Ashtekar variables. The states which we construct will be coherent states peaked over this point.
Consider first a single edge $l_i$ and thus one copy of $SU(2)$. 
There exist families $\phi^t_{l_i}\in L^2(SU(2))$ such that
$$ \lim_{t \to 0}\langle \phi^t_{l_i}, t \cl_{{\bf e}^a_i}\phi_{l_i}^t \rangle=2^{-2n}\mathrm{i}E_a^m(x_{j})\;,$$
and
$$\lim_{t \to 0}\langle \phi_{l_i}^t\otimes v, \nabla(l_i)\phi_{l_i}^t\otimes v \rangle=(v,h_{l_i}(A)v)\;,$$
where $v \in \bbC^2$, and $(,)$ denotes the inner product hereon; $x_{j}$ denotes the 'right' endpoint of $l_i$ (we assume that $l_i$ is oriented to the `right'), and the index ``$m$" in the $E^m_a$ refers to the direction of $l_i$. 
The factor $2^{-2n}$ is due to the fact that $\cl_{{\bf e}_j^a}$ corresponds to a flux operator with a surface determined by the lattice \cite{AGNP1}.
Corresponding statements hold for operators of the type 
$$f(\nabla(l_i))P(t \cl_{{\bf e}^1_i},t \cl_{{\bf e}^2_i},t \cl_{{\bf e}^3_i}),$$  
where $P$ is a polynomial in three variables, and $f$ is a smooth function on $SU(2)$, i.e.
$$ \lim_{t \to 0}\langle \phi^t_{l_i} f(\nabla(l_i))P(t \cl_{{\bf e}^1_i},t \cl_{{\bf e}^2_i},t \cl_{{\bf e}^3_i}) \phi^t_{l_i} \rangle=f(h_{l_i}(A))P(\mathrm{i}E_1^m,\mathrm{i}E_2^m,\mathrm{i}E_3^m)\;.$$
%
%
%
Finally define $\phi^t_n$ to be the product of all these states as a state in $L^2(\ca_{n})$. These states are essentially identical to the states constructed in \cite{TW} except that they are based on cubic lattices.
In the limit $n\rightarrow\infty$ these states produce the right semi-classical expectation value on all parallel transport operators in the infinite lattice. 

Note that the semi-classical state depends not only on a pair of Ashtekar variables but also on a spinor $v(x)$. \\

So far we have not specified exactly how the states $\phi^t_n$ are constructed. In \cite{BT1,TW,BT2} it is shown that different possibilities exist and that $\phi^t_n$ depends strongly on a choice of a so-called complexifier, which dictates exactly how the state is localized over the classical phase-space point. In this paper we shall not go into details on this otherwise crucial issue. Instead, we will take the following conditions, which are concerned with convergence of expectation values in the limit $n\rightarrow\infty$, as a definition of the state $\phi^t$ and simply conjecture the existence of the state. 

The conditions which we need are the following:  Let $fF$ be an element in $\mathbf{HD}(M) $ and let $\{\mathbbm{fF}_n\}$ be a corresponding sequence of approximations. Also, denote by $\hat{E}^m_a(x) =\{ t 2^{2n} \cl_{{\bf e}^a_{i_n}} \}$ a sequence of vector fields, where each vector field $\cl_{{\bf e}^a_{i_n}}$ is associated to the graph $\G_n$ and where $\{l_{i_n}\}$ is a sequence of edges in $\{\G_n\}$ whose end points converges towards $x$ in $M$ and where $"m"$ denote the direction of these edges (i.e. $x$, $y$ or $z$ directions). 

\begin{conj}
\label{concon}
There exist a choice of complexifier which gives a sequence $\phi^t = \{\phi^t_n\}$ of semi-classical states so that the following four requirements are satisfied:
\begin{eqnarray*}
&1. & \ce(fF):= \lim_{n\rightarrow\infty} \langle \phi^t_n \vert \varphi(\mathbbm{fF}_n)\vert \phi^t_n \rangle <    \infty
\\
&2.&  \ce(\hat{E}_m^a(x)):=    \lim_{n\rightarrow\infty} \langle \phi^t_n \vert    t 2^{2n}   \cl_{{\bf e}^a_{i_n}}^n   \vert \phi^t_n \rangle <    \infty
\\
&3.& \lim_{t\rightarrow 0} \ce(fF) = (v, \varphi_A(fF) v) \;.
\\
&4.& \lim_{t\rightarrow 0} \ce(\hat{E}^m_a(x)) = \mathrm{i} E^m_a(x)\;,
\end{eqnarray*}
\end{conj}
where 1. and 2. must hold independently of the choice of approximation and where we by $\varphi_A(fF)$ refer to the representation of $\mathbf{HD}(M) $ given by $A$. 
With these conditions satisfied $\phi^t$ is a state on $\mathbf{HD}(M) $, which gives a classical geometry in a semi-classical limit. All the following constructions depend on the existence of $\phi^t$.

\subsection{A $\mathbf{D}_{vol}(M) $-invariant semi-finite spectral triple}

Let us backpedal a little. At the finite level of a graph $\G_n$ we find that $$\phi^t_n\otimes \mathds{1}_{Cl}\otimes \mathds{1}_{2}\otimes \mathds{1}_{\vert {\bf v} \vert }$$ is a state on $\mathbf{HD}_n(M) $ where $ \mathds{1}_{Cl}$ is the identity element in the Clifford algebra $Cl(T^*\ca_n)$. For simplicity we shall also denote this state by $\phi^t_n$ and trust that no confusion will arise.

In fact, $\phi^t_n$ is a state on the larger algebra generated by $\mathbf{HD}_n(M) $ together with the spectral projections of $D_n$ and the diffeomorphisms in $\mathbf{D}_n(M) $. We denote this algebra $\mathbb{A}_n(M)$.

Alternatively, we can view $\phi^t_n$ as a "state" on $\mathbb{A}_n(M) $ which takes values in $\vert {\bf v} \vert \times \vert {\bf v} \vert$ matrices. To do this we evaluate the inner product in $H_n$ only with respect to integration over $\ca_n$ via the Haar measure and with respect to the trace over two-by-two matrices and with respect to the trace in the Clifford algebra $Cl(T^*\ca_n)$ (and {\it not} with respect to the trace over the vertices in $\G_n$ represented in $\vert {\bf v} \vert \times \vert {\bf v} \vert$ matrices). We choose the latter interpretation and apply the GNS construction to $\phi^t_n$ and denote the resulting Hilbert module by $H_{\phi^t_n}$. 
Note that $\mathbf{HD}_n(M) $, $\mathbf{D}_n(M) $ and $\mathbf{D}_{vol,n}(M) $ all act in $H_{\phi^t_n}$ as does the Dirac type operator $D_n$. Note also that $D_n$ commutes with the action of $\mathbf{D}_{vol,n}(M) $.

Now, denote by $\mathbf{D}'_{vol,n}(M) $ the commutant of $\mathbf{D}_{vol,n}(M) $ in $H_{\phi^t_n}$. Clearly $\mathbf{HD}_n(M) $ is not in $\mathbf{D}'_n(M) $ but the spectral projections of $D_n$ are.
The commutant will be too large an algebra to construct a spectral triple with and instead we will take the completion of the sub-algebra given by elements in $\mathbf{D}'_{vol,n}(M) $ which have a bounded commutator with $D_n$. This is the maximal possible choice, which we denote $\mathbb{X}_n(M)$.
We can thus view ${\phi^t_n}$ as a state which gives rise to a $(\mathbb{X}_n(M) ,\mathbf{D}_{vol,n}(M) )$ Hilbert bi-module. Furthermore, since $D_n$ commutes with $\mathbf{D}_{vol,n}(M) $ and since the commutator between $D_n$ and any element of $\mathbb{X}_{n}(M) $ is bounded we have in fact an unbounded Kasparov bi-module 
$$( \mathbb{X}_n(M)  ,  H_{\phi^t_n} ,D_n )$$
over $\mathbf{D}_{vol,n}(M) $. This Kasparov bi-module comes with an additional action of $\mathbf{HD}_n(M) $ and $\mathbf{D}_n(M) $. The commutator between $D_n$ and an element in $\mathbf{HD}_n(M) $ is bounded but of course $\mathbf{HD}_n(M) $ and $\mathbf{D}_{vol,n}(M) $ do not commute.

Thus we find that we have constructed an infinite sequence 
\begin{equation}
\{    (  \mathbb{X}_n(M)  ,  H_{\phi^t_n} ,D_n )   \} \qquad  n\in\mathbb{N}_+
\label{grantree}
\end{equation}
of unbounded Kasparov bi-modules associated to the infinite system $\OO$ of graphs in $M$. 

Notice that there is a natural trace over $\mathbf{D}_{vol,n}(M) $ given by the normalized matrix trace over $\vert {\bf v} \vert \times \vert {\bf v} \vert$ matrices as well as a trace over the Clifford algebra\footnote{in \cite{AGN1}-\cite{AGN3} the semi-finiteness of the spectral triple was with respect to a trace over the infinite dimensional Clifford algebra $Cl(T^*\ca)$.}. This turns (\ref{grantree}) into a sequence of semi-finite spectral triples.

The key question is what happens to these structures in the continuum limit $n\rightarrow\infty$. 
First of all, it is clear from our previous discussion that $\OO$ constitutes a coordinate system of $M$. 
Also, per construction we know that $\mathbf{HD}_n(M) $ and $\mathbf{D}_{vol,n}(M) $ converges into the $^*$-algebras $\mathbf{HD}(M) $ and $\mathbf{D}_{vol}(M) $ generated by holonomy-diffeomorphisms and volume-preserving diffeomorphisms of $M$. Finally, from conjecture \ref{concon} we have that $H_{\phi^t_n}$ converges into a $(\mathbb{X}(M) ,\mathbf{D}_{vol}(M) )$ bi-module with an additional action of $\mathbf{HD}(M) $. Here $\mathbb{X}(M)$ denote the limit of $\mathbb{X}_n(M)$. Thus, the key remaining question is what happens to the spectrum of $D_n$ as $n\rightarrow \infty$.

It is beyond the scope of this paper to analyze the existence of $D$ as an operator in $H_{\phi^t}$ and the behavior of its spectrum. Instead, we make the following two conjectures:
\begin{conj}
\label{conj1}
The infinite sequence $\{    ( \mathbb{X}_n(M)   ,  H_{\phi^t_n} ,D_n )   \} $ of Kasparov modules  assigned to $\OO$ gives a Kasparov module $\{    (  \mathbb{X}(M)  ,  H_{\phi^t} ,D )   \} $.
\end{conj}
\begin{conj}
\label{conj2}
The Kasparov module $\{    ( \mathbb{X}(M) ,  H_{\phi^t} ,D )   \} $ is a semi-finite spectral triple with respect to the trace over $Cl(T^*\ca)$  and the trace over $\mathbf{D}(M) $ induced by matrix traces over $\mathbf{D}_n(M)$.
\end{conj}

For these two conjectures to hold one must be careful to define the correct action of the algebra $\mathbf{D}_{vol,n}(M) $. First of all, one should define an action on the $\vert {\bf v} \vert \times \vert {\bf v} \vert$ matrices in $H_{\phi^t_n}$ as an algebra action build over the adjoint group action (and not, as we have done so far, as a pure left-action). In this way the diffeomorphisms will have an action on both copies of $M$ (or the vertices in $\G_n$) present. Second, as already mentioned one must have an additional action of $\mathbf{D}_{vol,n}(M) $ on the $L^2(\ca_n)$-part of the module $H_{\phi^t_n}$. This is necessary for $D$ to obtain a compact resolvent.

Notice that these conjectures are only concerned with the existence and spectral properties of the Dirac type operator $D$. The existence of a bi-module is guaranteed once we have the state $\phi^t=\{\phi^t_n\}$. The open question is whether $D$ exists as an operator in this limit and whether it is a Dirac type operator.

In the continuum limit  -- where we consider infinite sequences instead of finite approximations -- sums over points in $M$ become integrals. In particular, the trace over the $\vert {\bf v} \vert \times \vert {\bf v} \vert$ matrices turn into an integral
$$
\sum_{\small vertices}\rightarrow\int_M\
$$
which means that
 the matrix trace over $\mathbf{D}_{vol,n}(M)$ becomes an integral over those parts of $M$ on which a given element in $\mathbf{D}_{vol,n}(M)$ acts as the identity. We suspect that the spectrum of $D$ will likewise be continuous. 

So far everything has been built with respect to volume preserving diffeomorphisms $\mathbf{D}_{v}(M)$. It is not clear whether it is possible to formulate a spectral triple construction with respect to all the diffeomorphisms in $\mathbf{D}(M)$. Here the critical question is whether one can formulate an action of $\mathbf{D}_{vol,n}(M)$ on $L^2(\ca)$ that commutes with $D_n$. Since there is a natural action of diffeomorphisms on the full configuration space $\ca$ one might expect such an action to exist.

In the earlier papers \cite{AGN1}-\cite{AGN3} we constructed a semi-finite spectral triple associated to a projective system of graphs. In this case the semi-finiteness was with respect to a trace over the Clifford algebra $Cl(T^*\ca)$.\\

We find it remarkable that the application of noncommutative geometry -- Dirac type operators and spectral triples --  is somewhat forced upon us by this issue of diffeomorphism invariance. The Dirac type operator $D$ is one of the simplest operators that involve the quantized triad operators and which is invariant under volume-preserving diffeomorphisms. \\

A spectral triple is useful since it gives us access to well defined quantities, for instance via the spectral action \cite{Chamseddine:1996rw,Chamseddine:1996zu}. 
To have a semi-finite spectral triple where the semi-finiteness is with respect to an algebra generated by local volume-preserving diffeomorphisms is highly desirable since such a construction guarantees us that the objects we build from our basic building blocks -- the Dirac type operator and the algebra -- will be invariant with respect to these diffeomorphisms. Therefore it is also important to determine whether $D$ might in fact commute with {\it all} spatial diffeomorphisms. We are, however, intrigued by the possibility that our construction singles out the volume-preserving diffeomorphisms and leaves open the possibility of having scale-dependent objects.






\section{Emergent QFT and almost commutative geometry}

The construction of a semi-classical state 
obviously points in the direction of a semi-classical approximation. This section, which is partly based on previously published results, is concerned with the semi-classical emergence of i) key elements of (fermionic) quantum field theory and ii) an almost commutative geometry involving $C^{\infty}(M)$. We recall that the latter type of algebra is the key ingredient in the work by Connes and coworkers on the standard model.

\subsection{Connection to fermionic quantum field theory}

In the papers \cite{AGNP1}-\cite{Aastrup:2011dt} we showed that key elements of fermionic quantum field theory emerge in a semi-classical approximation from the spectral triple construction which we first presented in \cite{AGN1,AGN2} and further developed in \cite{AGN3}. This construction is based on a projective system of nested cubic lattices and is, at the level of a finite lattice, essentially identical to the construction presented in section \ref{hulk}. 
The key difference between the construction analyzed in \cite{AGNP1}-\cite{Aastrup:2011dt} and the construction presented in this paper is the way the continuum limit is performed. 

In \cite{AGNP1}-\cite{Aastrup:2011dt} the Hilbert space is an inductive limit of intermediate Hilbert spaces associated to finite graphs. From this Hilbert space we found an infinite sequence of semi-classical states. Whereas each element in this sequence belong to the Hilbert space the limit does not. In the present paper this is different: here states on $\mathbf{HD}(M) $ are exactly those infinite sequences of intermediate states associated to finite graphs. 

In \cite{AGNP1}-\cite{Aastrup:2011dt} we considered the double limit where we first perform a semi-classical approximation at a finite level and second take a continuum limit. We found that the Dirac Hamiltonian on $M$ emerges in this limit from the expectation value of a Dirac type operator on these semi-classical states, see \cite{Aastrup:2010ds}, as does many-particle states and a Fock space structure, see \cite{Aastrup:2011dt}. 

In this paper the order of this double limit is effectively reversed while most of the constructions -- the Dirac operator and semi-classical states -- are kept essentially unaltered. This means that the results on emergent fermionic quantum field theory will also hold here, possible with minor modifications. 
Since we have nothing substantially new to add to this part of the story we simply give a brief outline of the basic mechanism and refer the reader to the paper \cite{Aastrup:2011dt} for details.\\

Before we continue we will change the representation of the holonomy-diffeomorphisms algebra. Instead of the left-action on $H_n$ and $H_{\phi^t}$ we will in the following assume that we have an algebra representation built over a conjugate group action. This is a technical detail and at the present level of analysis we are not certain whether we need it or not. But since it makes the subsequent analysis easier we adopt it here.

The first step is to find states on which the Dirac type operator $D$ has a non-vanishing expectation value. Since $D$ is odd with respect to the Clifford algebra the state must be a sum of both even and odd parts. Let us consider the simplest case where we have a state $\xi$ which is a sum of two parts $\xi= \xi_1 + \xi_2$, where $\xi_1$ a scalar in the Clifford algebra and $\xi_2$ includes a single element hereof.

We consider first a finite level $\G_n$. An element of the Clifford algebra can appear in $H_{\phi^t}$ via a commutator (we omit the constant $\a_n$)
\begin{equation}
[D,\nabla(l_i)] = e_i^a \sigma_a \nabla(l_i)  
\nn
\end{equation}
that acts on $\phi^t$. This should be understood as a single entry in a $\vert {\bf v} \vert \times \vert {\bf v} \vert$ matrix, which is one step off the diagonal corresponding to the start point of the edge $l_i$.  There is an additional degree of freedom stemming from the $2\times 2$ matrix factor in $H_{\phi^t}$, which permit us to add a $2\times 2$ matrix that we denote $\psi(v_i)$. Here $v_i$ denotes the vertex where $l_i$ ends. In total we get a term
$$
\xi_2=e_i^a \sigma_a \nabla(l_i) \psi(v_i)  \nabla(l_i^*)  \phi^t_n  \;,
$$
which is still to be understood as an entry in a $\vert {\bf v} \vert \times \vert {\bf v} \vert$ matrix. Here we have used that we have a conjugate action of the holonomies.

Now, the other part of the state has to be a scalar wrt the Clifford algebra and we choose one of the form
$$
\xi_1=P_{v_j}\psi(v_j) \phi^t_n
$$
where $P_{v_j}$ is a projection on the vertex $v_j$ and $\psi(v_j)$ is another $2\times 2$ matrix. Finally, we see that in order for $D$ to have a non-vanishing expectation value we must let $v_j$ coincide with the endpoint of $l_i$. 

The key mechanism is now that when we take the expectation value of the Dirac type operator $D$ on such a state then the trace over the Clifford algebra will couple the Lie algebra index on $ \cl_{{\bf e}_i^a} $ (coming from $D$) and $\sigma_a$ (coming from $\xi_2$). Since the expectation value of  $\cl_{{\bf e}_i^a}$ on a coherent states gives a densitized triad field $E_m^a$ (we omit scaling factors) we get from these two parts the contraction
$$
\sim E_m^a \sigma_a
$$
where $m$ refers to the direction of $l_i$. Furthermore, from $\nabla(l_i) \psi(v_i)\nabla(l_i^*)$ we get
$$
\sim \pa_m \psi(v_j) + [A_m,\psi(v_j)]
$$
where we have already anticipated the continuum limit. 

We now sum up the $\xi$'s associated to all vertices in $\G_n$ and denote the sum $\X^t$.   Ignoring minor technicalities we get
$$
\lim_{t\rightarrow 0}\langle \X^t\vert D\vert \X^t\rangle =\int_M \psi \not\hspace{-1mm} D \psi
$$
with $\not\hspace{-2mm} D= E_m^a \sigma_a( \pa_m + [A_m,\cdot])$. Here we have ignored the crucial issue of determining the various scaling factors, see \cite{Aastrup:2011dt}.

To get the Dirac Hamiltonian one can perform a local transformation of the $2\times 2$ factor in $\X^t$
$$
\sigma_a\rightarrow M_i \sigma_a
$$
where $M_i$ is a self-adjoint $2\times 2$ matrix associated to the vertex $v_i$. The lapse and shift fields arise from $M_i = N \mathds{1}_2 + \sigma_a N^a$ and gives us
$$
\int_M \psi \not\hspace{-1mm} D \psi \rightarrow \int_M \psi \left( N \not\hspace{-1mm} D + N^a \pa_a\right) \psi\;,
$$
which is the Dirac Hamiltonian (see \cite{Aastrup:2009dy} for an early version of this idea).

Many particle states then arise from states with a more complicated Clifford algebra structure, see \cite{Aastrup:2011dt}.

Notice that effectively it is the Clifford algebra in the semi-classical approximation that provide the construction with spatial derivations. 

Also, we find it intriguing that the CAR algebra and the Fock space structure in this framework are so intimately quantum gravitational in their origin. 


\subsection{An emergent almost commutative geometry}

We are now going to consider the semi-classical limit of the holonomy-diffeomorphism algebra.
Recall that the algebra $\mathbf{HD}(M) $ can be formulated as the closure of a semi-direct product
$$
\mathbf{HD}(M)  = C^\infty(M)  \rtimes  \cF  / I\;.
$$
We are interested in the the semi-classical approximation $t\rightarrow 0$ of the GNS representation of $\mathbf{HD}(M) $ in $H_{\phi^t} $. In this limit $\mathbf{HD}(M) $ reduces to the algebra
\begin{equation}
 \left(C^\infty(M)\otimes M_2(\mathbb{C})\right) \rtimes \mbox{Diff}(M)\;,
\label{almost}
\end{equation}
where $\mbox{Diff}(M)$ is the group of diffeomorphisms on $M$. This is so because the holonomies on a fixed classical geometry generate a two-by-two matrix algebra\footnote{We assume we are considering a semi-classical analysis around a irreducible connection.}.
Thus we find the almost commutative algebra $C^\infty(M)\otimes M_2(\mathbb{C}) $ as a subalgebra.

In fact there is a sub-algebra of $ \mathbf{HD}(M)$, which involve those flows that preserve the points in $M$. For a given representation of  $ \mathbf{HD}(M)$ this sub-algebra has the form $C^\infty(M)\otimes M_2(\mathbb{C}) $.

Notice also that the space $L^2(M,S)$ -- where $S$ refers to two-spinors on $M$ -- emerge from the GNS construction of $\phi^t$ in the semi-classical limit. This limit simply gives a Riemann integral from the trace over the $\vert {\bf v} \vert \times \vert {\bf v} \vert $ matrices.

Finally, we note that the spatial Dirac operator $\not\hspace{-1.9mm}D$, which emerge in a semi-classical approximation, interact not only with the smooth function on $M$ but also with the matrix factor $M_2(\mathbb{C})$. The exact nature of this interaction still remains to be worked out, but it is clear that the commutator between the Dirac type operator (\ref{DIRAC}) and the holonomy-diffeomorphism algebra is non-zero outside the classical limit and will, therefore, give some kind of interaction in the limit.

Thus, a picture of an emergent spectral triple over an almost commutative algebra seems to arise from our construction. 
A couple of comments are in order:
\begin{itemize}
\item[-]
first of all let us stress that we are in no position to explain the particular structure of the almost commutative algebra used by Connes and coworkers. A more direct assessment of whether the emergent algebra in (\ref{almost}) might be related to Connes work is hindered by the fact that the latter operates with a 4-dimensional Lagrangian picture whereas the former is formulated in a 3-dimensional Hamiltonian picture. It seems that a reformulation of Connes work in a Hamiltonian picture would be needed to be able to compare the two.
\item[-]
another issue is that the emergent matrix factor $M_2(\mathbb{C})$ acts on the spinors in $L^2(M,S)$ and not on an additional finite-dimensional Hilbert space as it is the case in Connes' work on the standard model.
\end{itemize}

\section{The dynamics of quantum gravity}

A key question, which remains open, is how the dynamics of quantum gravity fits into the machinery presented in this paper. Although we do not have an answer to this question we will, in this section, show how a quantization of the constraint system (\ref{constraints}) can be obtained by a certain arrangement of the Dirac type operator $D$ and elements of the algebra $\mathbf{HD}(M) $. There are several ways to accomplish this and since we do not yet see a fundamental principle behind this arrangement we intend this section simply as a suggestion.\\

Consider first a sequence $\{\mathbbm{fF}_{n}\}$ which approximates an infinitesimal element $fF_{ds}\in \mathbf{HD}(M) $. The commutator between this element and the Dirac type operator $D$
$$
 [D,fF_{ds}] = \{ [D_n,\mathbbm{fF}_{n}] \}
$$
amounts to a one-form in the sense of noncommutative geometry \cite{ConnesBook}. 
The sequence $\{\mathbbm{fF}_{n}\}$ is to be understood as an equivalence class, and it is possible to choose an element in this class where each $\mathbbm{fF}_{n}$ involve only paths which parallel transport along single edges. Thus, the only non-zero entries in $\mathbbm{fF}_{n}$ can be chosen to be entries which connect adjacent vertices. With this  convention in place let $v_i,v_j$ be adjacent vertices and compute the $(v_i,v_j)$ entry of $ [D_n,\mathbbm{fF}_{n}] $
$$
 [D_n,\mathbbm{fF}_{n}] (v_i,v_j) = 
e^a_j \sigma^a \nabla(l_k) (v_i,v_j)
$$
where $\nabla(l_k)$ is an element in the copy of $G$ assigned to the $k$'s edge connecting $v_i$ and $v_j$.
We now introduce the fluctuated Dirac type operator
$$
\tilde{D} := D +  [D,fF_{ds}]   =\{D_n + [D_n, \mathbbm{fF}_{n} ] \}
$$
and compute its square
\begin{equation}
\tilde{D}_n^2 = D_n^2 + [D_n,  [D_n, \mathbbm{fF}_{n} ]   ]_+ +  ([D_n, \mathbbm{fF}_{n} ] )^2 \;,
\label{Dtilde2}
\nn
\end{equation}
where $[\cdot,\cdot]_+$ denotes the anti-commutator. Consider the square of the second term of this expression and let us only consider the terms which are scalar with respect to the Clifford bundle. 
We find
\begin{equation}
( [D_n, [D_n, \mathbbm{fF}_{n} ]   ]_+)^2 = [ d_{e_i^a},\nabla(l_i)]_+  \sigma^a [ d_{e_j^b},\nabla(l_j)]_+  \sigma^b \;.
\label{HAMI}
\end{equation}
Such a term will only give a non-zero expectation value in $H_{\phi^t_n}$ whenever $l_i$ and $l_j$ are connected, and when the term is evaluated on a state which produces additional edges $l_k$ and $l_l$ such that the path
$$
\{ l_i,l_j,l_k,l_l \}
$$
forms a (possible trivial) loop. 
In that case the expectation value of a sequence of such operators will lead to a classical expression of the form
\begin{equation}
\int_M  N \e^{ab}_{\;\; \;c}  E^{m}_a  E^{n}_b  F^c_{mn}\;.
\label{equation...}
\nn
\end{equation}
which is exactly the form of the Hamilton constraint formulated in terms of Ashtekar variables (\ref{constraints}), where the lapse field $N$ arise from a local weight of the state.

Several comments are in order:
\begin{itemize}
\item[-]
we have in these considerations ignored the important issue of scaling factors. The semi-classical states involve scaling factors, which must be countered in an operator in order to render finite and non-vanishing expectation values. Also, issues concerning the volume element $\sqrt{g}$ are likewise swept under the rug.
\item[-]
it is possible to modify the operator (\ref{HAMI}) to give also the diffeomorphism constraint. Such a modification is given by a local transformation of the two-by-two factor in $H_{\phi^t_n}$.
\item[-]
the operator (\ref{HAMI}) is built from a two-form (two commutators) and two Dirac type operators. If we imagine that the two-form is provided by the state -- as was the case in the previous section --, then this operator is in effect quadratic in $D$.
\item[-]
here we have simply put in an infinitesimal object $fF_{ds}$ by hand. From the point of view of our construction this is not a natural object to consider. Rather, we believe that such an element should be singled out by the Clifford algebra $Cl(T^*\ca)$ much alike the mechanism behind the emergence of the Dirac Hamiltonian.

\end{itemize}

\section{Comparison to loop quantum gravity}

The approach to quantum gravity presented in this paper is in some respects similar to loop quantum gravity \cite{AL1,Thiemann:2001yy,Sahlmann:2010zf} but differs decisively on several critical points. In this section we will point out the main similarities and differences between the two.\\

LQG takes its point of departure with a phase space of Ashtekar variables \cite{Ashtekar:1986yd,Ashtekar:1987gu}. The key technical tool is a choice of an algebra $\cb$ generated by parallel transports\footnote{the actual choice of algebra in LQG varies somewhat depending on whether one considers earlier or later versions. These variations are not important for our discussion here.} restricted to a projective system of piece-wise analytic graphs in $\S$. The spectrum of $\cb$ contains the space of smooth connections as a dense subset. This result, due to Ashtekar and Lewandowski \cite{Ashtekar:1993wf}, permits the construction of a kinematical Hilbert space $H_{\tiny kin}$ as an inductive limit of intermediate Hilbert spaces associated to finite graphs. The resulting measure on the spectrum of $\cb$ is an inductive limit of Haar measures on various copies of $SU(2)$.
 The kinematical Hilbert space caries a representation of the Poisson algebra and is the starting point of the LQG-quantization where the constraint algebra is realized as operators in $H_{\tiny kin}$.

The non-separability of $H_{\tiny kin}$ is a direct consequence of the choice of piece-wise analytic graphs. This choice, in turn, stems from the wish to have an action of the diffeomorphism group acting in $H_{\tiny kin}$.

As already mentioned, the spectrum of the algebra $\cb$ contains the space of smooth connections as a dense subset. This subset has, however, zero measure with respect to the Ashtekar-Lewandowski measure. The bulk of the spectrum is given by so-called generalized connections which are non-smooth and non-measurable with respect to $L^2(M)$. An example of a generalized connection is an object which has support on a finite graph only.

An important consequence of constructing the kinematical Hilbert space as an inductive limit is that this requires the gauge group to be compact. This restriction seriously complicates the LQG-quantization program due to its impact on the constraint algebra (see \cite{Sahlmann:2010zf} for details).\\

Thus, both LQG and the approach presented in this paper start with the phase space of Ashtekar variables and uses parallel transports as a key ingredient.  Also, both approaches use graphs, albeit in different ways.
The main {\it differences} between the two approaches are:
\begin{itemize}
\item[-]
the choice of algebra. The choice of $\mathbf{HD}(M) $ made in this paper entails a number of fundamental differences to LQG. The most important one is the way an action of the diffeomorphism group is {built into} the construction. 
\item[-]
the spectrum of $\mathbf{HD}(M) $ does not contain the generalized connections found in the LQG spectrum.
\item[-]
the construction of the kinematical Hilbert space is completely different in the two approaches. Whereas LQG operates with a universal kinematical Hilbert space we suggest to consider a Hilbert space generated by a state on $\mathbf{HD}(M) $. The natural choice -- and the only one which we are able to device a method of construction -- is a semi-classical state. This means that each semi-classical approximation comes with a {\it different} kinematical Hilbert space.
\item[-]
this entails a different measure on the configuration space of connections. The alternative to the Ashtekar-Lewandowski measure is provided by the GNS construction.
\item[-]
the use of graphs in this paper is not that of a projective-inductive limit. Rather, it is the formulation of continuum objects in terms of infinite sequences of cubic graphs, which represent a coordinate system. 
\item[-]
the constructions presented in this paper do not impose the same restrictions on the topology of the structure group as are encountered in LQG.
\item[-]
matter couplings arise naturally in the constructions presented here whereas they must be introduced by hand in LQG. Also, we find a possible connection to Connes work on the Standard Model.
\end{itemize}

Another way of comparing the two approaches is to note that LQG can -- in a certain sense -- be viewed as a discretized version of the construction presented in this paper. In \cite{AGnew} we show that the representation theory of the discretized holonomy-diffeomorphism algebra generated by 
$$
 C_d(M)  \rtimes  \cF  / I\;,
$$
where $ C_d(M)$ is the algebra of functions on $M$ with finite support, is characterized by generalized connections.
This means that the algebra spectrum in LQG arise from a discretization of our construction. Furthermore, in \cite{AGnew} we show that generalized connections do {\it not} form a representation of the holonomy-diffeomorphism algebra.

\section{Discussion}
\label{Discussion}

In this paper we have presented a new approach to a unified theory of quantum gravity, which combines canonical quantum gravity with elements of noncommutative geometry.

The two sets of variables, which in this approach capture the kinematics of quantum gravity, have basic geometric origins: the holonomy-diffeomorphism algebra tell us how {\it things} are moved around in $M$, and the Dirac type operator can -- due to its semi-classical behavior -- be interpreted as a quantization of a spatial Dirac operator on $M$, which carries metric information hereof. These quantum variables are organized in a spectral triple construction.

A major insight gained in this paper is that this application of noncommutative geometry is strongly motivated by the issue of diffeomorphism invariance. Once we have settled on the holonomy-diffeomorphism algebra the Dirac type operator is, due to its invariance properties, the natural choice for an operator containing the conjugate variables.

The result that the separable part of the spectrum of the holonomy-diffeomorphism algebra is a space of measurable connections then further solidifies the interpretation of our construction as a framework of canonical quantum gravity based on Ashtekar connections.

A semi-classical analysis -- carried out in \cite{AGNP1}-\cite{AGN3} but applicable for our construction -- shows that elements of quantum field theory emerge from the spectral triple construction in a semi-classical limit. This is a strong indication that the spectral triple construction naturally incorporates quantized matter fields. This together with
the observation that an almost commutative algebra -- and possible an almost commutative geometry -- emerge in the semi-classical limit further indicates that the spectral triple is a framework of unification.

All together a picture emerges of a non-perturbative theory of quantum gravity from which elements of quantum field theory and unification arise in a semi-classical approximation.\\

The most important question that now needs to be addressed is that of convergence. Here, two issues are crucial: the existence of the semi-classical state and the existence and spectral properties of the Dirac type operator. Since elements of quantum field theory emerge from our construction we find it plausible that this issue will entail subtleties corresponding to renormalization theory. At the present state of the project the investigation of this issue is certainly within reach.

We are at the moment unable to completely determine the spectrum of $\mathbf{HD}(M) $. We know that its separable part is given by measurable connections and that its non-separable part does {\it not} include the generalized connections found in LQG. It is, however, an open question what other non-separable elements it may contain. We suspect that the reason why we are unable to completely rule out any non-separable elements is that we have failed to choose the right topology for $\mathbf{HD}(M) $. The $C^*$-topology, which we have employed in this paper, is partly insensitive to the topology of $M$. It would be more natural to work with the strong topology, which would then lead to a von Neumann algebra setting. Also, with the strong topology we would ensure that the algebra can be approximated by lattices -- something which is in fact not the case with the $C^*$-topology.

In fact, the choice of the correct algebra topology seems to be a crucial issue. Since the strong topology is given by a Hilbert space representation it seems natural that the holonomy-diffeomorphism algebra should be defined as a von Neumann algebra with respect to the strong topology given by the GNS construction of a semi-classical state. The GNS construction gives what amounts to a kinematical Hilbert space, which is the only available Hilbert space in our approach. It seems likely that such a von Neumann algebra would be independent of which state one uses since different semiclassical states can be considered as different "Gaussian peaks" on the "space of connections", and therefore belong to the same measure class. The key issue would then be to show that such an algebra is independent of the lattices/coordinate system, which is needed to construct the semi-classical states. 

Another central question is to determine how a dynamical principle of quantum gravity fits into the construction. Operators, which correspond to the classical constraint system of gravity formulated in terms of Ashtekar variables, must be identified. Of course, one possibility is simply to write down an operator, which has the correct classical limit. This is possible and one can even write down such an operator in terms of the Dirac type operator and elements of $\mathbf{HD}(M) $. Such an approach will, however, always be marred by ambiguities and lack of canonicity. A more appealing possibility is to seek a dynamical principle within the mathematical machinery of noncommutative geometry. In particular, the theory of Tomita and Takesaki states that given a cyclic and separating state on a von Neumann algebra there exist a canonical time flow in the form of a one-parameter group of automorphisms. If we consider the algebra generated by $\mathbf{HD}(M) $ and spectral projections of the Dirac type operator, then the semi-classical state will, provided it is separating, generate such a flow. This would imply that the dynamics of quantum gravity is state dependent\footnote{up to inner equivalence.} - an idea already considered in \cite{Connes:1994hv} and \cite{Bertozzini:2010us}. Since Tomita-Takesaki theory deals with von Neumann algebras it will also for this purpose be important to select the correct algebra topology.

The emergence of an almost commutative algebra in the semi-classical limit is, as already mentioned, a strong indication that we are indeed dealing with a theory of unification. In the work by Connes and co-workers on the standard model, it is the inner automorphisms of such an algebra, which generate the entire bosonic sector of the standard model. Since the construction presented in this paper is formulated in the Hamiltonian picture, a direct comparison to Connes work is not straight forward. It is therefore important to first clarify to what extend an almost commutative geometry does emerge -- and this again involves the question of convergence -- and second to find a way to compare it to an almost commutative geometry in the Lagrangian picture. Some refinements of our construction -- for instance by introducing a real structure -- should be made and will be important for this issue.


We believe that a theory of quantum gravity based on parallel transport of the Ashtekar connection {\it should} involve fermionic degrees of freedom in an intrinsic manner. After all, the Ashtekar connection does involve the spin-connection and its parallel transport acts naturally on spinors. It is therefore, to our minds, very reassuring that fermionic degrees of freedom do emerge from our construction in a semi-classical analysis.
A number of questions concerning this emergence still remains to be clarified -- see \cite{Aastrup:2011dt} for a thorough discussion --, but perhaps the most interesting question is whether and how elements of renormalization theory might emerge. This question is clearly linked to the study of the spectral properties of $D$ and the existence of the semi-classical state $\phi^t$. 
We find the emergent relationship between quantum gravity and quantum field theory rather elegant: That (fermionic) quantum field theory should be understood as the low-energy limit of quantum gravity and that the Fock space structure emerge from a labeling (via the Clifford algebra $Cl(T^*\ca)$) of quantum gravitational degrees of freedom. 

The result that the Dirac type operator $D$ commutes with spatial volume-preserving diffeomorphisms -- but perhaps not all spatial diffeomorphisms -- raises a number of questions. First of all, it is necessary to determine whether $D$ actually commutes with {\it all} spatial diffeomorphisms. The analysis so far depend on observations made at a finite approximation. It may be that these observations do not hold in the continuum limit. Also, it might be possible to change the lattice approximation so that $D_n$ commutes with all diffeomorphisms restricted to a given level of approximation. 
Alternatively, it may very well be that the Dirac type operator -- and therewith the entire construction -- is stuck with a certain amount of background dependency. In particular, it seems that the construction has a built-in scale dependency in the sense that it is not invariant under arbitrary changes to the volume element. We suspect that this again might hint at a connection to renormalization theory, which also operates with a natural scale dependency. This issue is closely connected to the $\sqrt{g}$-issue encountered in \cite{AGNP1}-\cite{Aastrup:2011dt}. There the emergence of the Dirac Hamiltonian in a semi-classical approximation was studied and it was found that the normalization of spinors came with a similar scale dependency.

As already mentioned, once the toolbox of noncommutative geometry is opened a natural step is to introduce a real structure and a right-action of the algebra. For an ordinary spin-manifold the real structure is given by charge conjugation. In the setup presented in this paper it should be given by invertion of the holonomy-diffeomorphisms together with complex conjugation of the functions. We suspect that this issue is related to  another open issue: That the original Ashtekar connection is a {\it complexified} $SU(2)$ connection. We think that the natural way to obtain this complexification is via a real structure.

Hidden within the two issues concerning of the dynamics and the complexified $SU(2)$ connections lurks a very intriguing question. If it is possible to derive the dynamics of quantum gravity from the spectral triple construction -- for instance via Tomita Takesaki theory --  then it should be possible to {\it read off} the space-time signature (Lorentzian vs. Euclidean) from the derived dynamics, for instance from a moduli operator.

\subsection*{Acknowledgements}

We would like to thank the Mathematical Sciences Institute at ANU, Australia, and the Mathematics Department at Caltech, California, for offering kind hospitality during two visits. We are thankful to M. Marcolli, R. Nest and M. Paschke for useful discussions. The second author thanks IHES, France, for kind hospitality during a visit. The second author would like to thank A. Warburton. 

\end{document}